# A calculation method of nuclear cross-sections of therapeutic proton beams by the collective model and elements of the extended nuclear-shell theory

W. Ulmer[1,2,3] and E. Matsinos[1]

[1]Varian Medical Systems, Baden, Switzerland, [2]MPI of Physics, Göttingen, Germany, [3]Klinikum Frankfurt/Oder, Germany

E-Mail: waldemar.ulmer@gmx.net

**Abstract**: An analysis of total nuclear cross-sections of various nuclei is presented, which yields detailed knowledge on the different physical processes such as potential/resonance scatter and nuclear reactions. The physical base for potential/resonance scatter and the threshold energy resulting from Coulomb repulsion of nuclei are collective/oscillator models. The part pertaining to the nuclear reactions can only be determined by the microscopic theory (Schrödinger equation and strong interactions). The physical impact is the fluence decrease of proton beams in different media, the scatter behavior of secondary particles, and a 'translation' of the results of the microscopic theory to the collective model.

## 1. Introduction

The accurate knowledge of the total nuclear cross-section $Q^{tot}$ resulting from proton – nuclei interactions is a decisive feature of proton-therapy planning with advanced models and Monte-Carlo calculations. Nuclear cross-sections determine the fluence decrease of primary protons, different ranges and scatter angles of secondary particles (mainly protons), collimator scatter and passage of protons through bones, implants, etc., and the creation of heavy recoils, which usually undergo $\beta^+$ decay and emission of $\gamma$ quanta. Since water represents the usual reference medium for the measurement/calculation of Bragg curves, we consider, at first, the behavior of $Q^{tot}$ of the oxygen nucleus.

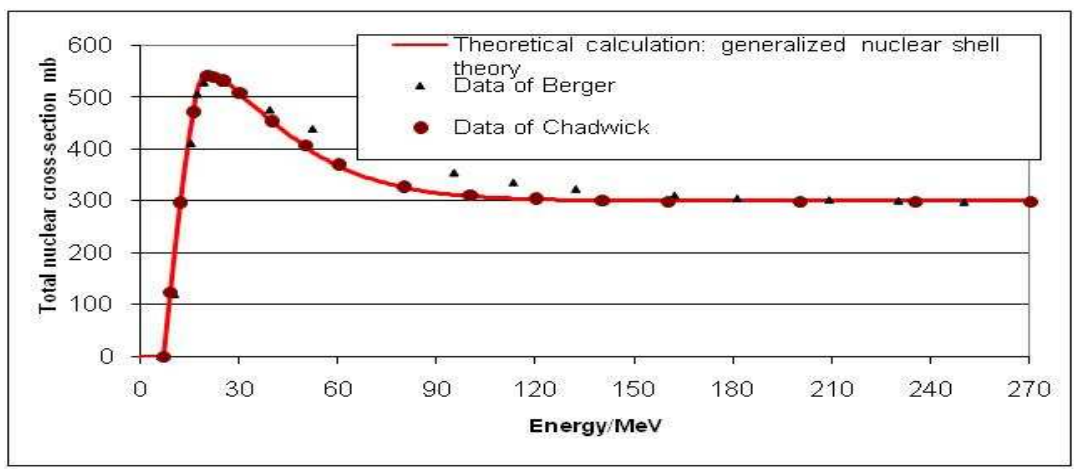

**Figure 1.** Total nuclear cross-section Q $^{tot}$ of oxygen (Berger et al 1993, and 2000, Chadwick et al 1996, Ulmer 2007)

Figure 1 shows that, for protons, a threshold energy $E_{Th}$ = 7 MeV exists to surmount the Coulomb repulsion of oxygen. At E = $E_{res}$ = 20.12 MeV, $Q^{tot}$ exhibits a resonance maximum and a Gaussian shape in the environment. For E > 50 MeV, Q decreases exponentially; at E ≈ 120 MeV, the asymptotic behavior is reached. Using a method of Segrè (1964) the fluence decrease of primary protons $\Phi_{pp}$ can be calculated (Ulmer 2007):

$$\left. \begin{array}{l} \Phi_{pp} = \frac{1}{2}[erf((R_{CSDA}-z)/\tau)+1] \cdot [1-uq\,\frac{z}{R_{CSDA}}] \\[2mm] uq = (E_0 - E_{Th})/Mc^2)^f\,;\ \ E_{Th} = 7\,MeV\,(oxygen);\ Mc^2 = 938.27\ MeV;\ f = 1.032 \end{array} \right\} \ (1)$$

Please note that the power *f = 1.032* is only valid for oxygen. Formula (1) with different *f* is also valid for other media; it will be stated in Eq. (3). It might be surprising that in formula (1) the error function *erf* and the rms-value τ of a Gaussian appear. In principle, the behavior of $\Phi_{pp}$, valid within the CSDA-framework, should be a straight line as long as E = $E_{Th}$ is not yet reached. For E < $E_{Th}$ to E = 0, $\Phi_{pp}$ should be constant; at E = 0 (z = $R_{CSDA}$) a jump to $\Phi_{pp}$ = 0 is expected. However, due to energy/range straggling, the proton beam can never remain mono-energetic in the sense of CSDA. Since τ refers to the half-width of a Gaussian convolution, we introduce 'roundness' in the shape. The range of 7 MeV protons is less than 1 mm; therefore, we cannot verify whether $\Phi_{pp}$ is constant in the fluence profile of primary protons. The definition of the half-width parameter τ (energy/range straggling) is not a subject of this paper (see e.g. Ulmer 2007).

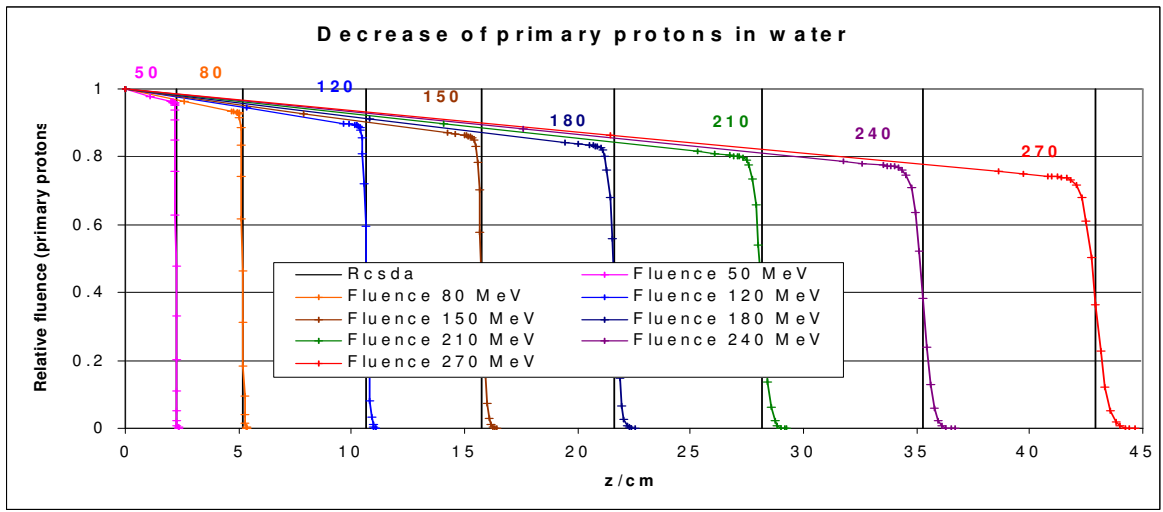

**Figure 2.** Decrease of the proton fluence in water due to nuclear interactions according to Figure 1.

Since we shall verify that we have to distinguish between two kinds of secondary protons, the determination of the fluence of secondary protons (sp) cannot be accounted for with the help of formula (1). A principal feature of this communication is the range $R_{strong}$ of the strong interaction in dependence of the nuclear mass unit A:

$$R_{strong} = A^{1/3} \cdot 1.2 \cdot 10^{-13} \ cm \quad (2)$$

According to Eq. (1) a determination of the power function $f$ is required to calculate the proton fluence decrease in rather different media:

$$f(Z, A) = a \cdot A^{-1} + b \cdot A^{-2/3} + c \cdot A^{-1/3} + d \cdot Z \cdot A^{-1/2} \quad (3)$$

A correspondence between the powers of $A$, inclusive their coefficients $a - d$ (Table 1) according to Eq. (3), and the essential underlying physical processes will be given in section 2.2. The goal of this communication is to determine some features of the $Q^{tot}$ in terms of the nuclear charge number Z, the nuclear mass unit A and/or $R_{Strong}$ by the toolkits of the collective model and extended nuclear-shell theory. In order to gain the necessary simplicity the latter method is presented in Appendix A.

**Table 1.** Parameters of the power function $f$ (Z, A).

| a | b | c | d |
|---|---|---|---|
| -0.087660001 | -6.379250217 | 5.4014005 | -0.05427999 |

Since Appendices A and B are only of interest for a small group of radiation physicists, we have made them available only via the online version.

## 2. Methods

Measurement data of $Q^{tot}$ for various nuclei have been made available by the Scientific Los Alamos Library; we refer to these measurements as the Los Alamos data (*LAD*). An analysis of part of these data has been published (Chadwick et al 1996, Ulmer 2007, Ulmer et al 2009).

### 2.1 Threshold energy $E_{Th}$

The threshold energy $E_{Th}$ sets the effective Coulomb repulsion energy between proton and nucleus Z. Since the nucleons are not localized at one point in space, the effective interaction charge with the external proton has to be determined. An analysis of the *LAD* suggests the following connection for nuclei with Z ≤ 40:

$$\left.\begin{array}{l} Z \geq 6,\ \kappa = 1.659: \\ E_{Th} = C_Z \cdot Z^\kappa \cdot F \\ F = (A/2Z) \cdot (a_0 + a_1/A + a_2/A^{p1} + a_3/A^{p2} + a_4/A^{p3} \end{array}\right\} (4)$$

$$\left.\begin{array}{l} Z < 6: \\ E_{Th} = C_Z \cdot Z^\kappa \cdot (6 - Z) \cdot F \\ \kappa = 1.659 + 0.341 \cdot (6 - Z)/5 \\ F = F(C_6^{12}) \end{array}\right\} (5)$$

The proportionality factor $C_Z$ is given by:

$$C_Z = 0.222265\ MeV \qquad (6)$$

Function *F(A, Z)* represents a form factor function for the inner structure of a nucleus due to nucleon – nucleon interactions, which accounts for that only a small percentage of protons become effective at surface of a nucleus.

**Table 2.** Parameters of the form factor function *F(A, Z)* according to Eq. (4).

| a₀ | a₁ | a₂ | a₃ | a₄ | p1 | p2 | p3 |
|---|---|---|---|---|---|---|---|
| 2.1726 | -335.0440 | 479.5400 | -194.9400 | 11.7125 | 0.76965 | 0.5575 | 0.3405 |

Formula (4) is only applicable, if 2·Z ≈ A, i.e. proton number ≈ neutron number. Nuclei with Z < 6 are corrected by Eq. (5), since they do not satisfy rotation invariance. Relation (4) is founded theoretically and is improved by the harmonic-oscillator model (Ulmer et al 2009). At r = $R_{Strong}$ and for isoscalar nuclei with A = 2·Z, the 3D harmonic-oscillator model of nuclear-shell theory assumes the shape (section 2.3 and Appendix A):

$$(-U_0 + \tfrac{1}{2}M \cdot A \cdot \omega_0^2 \cdot R_{Strong}^2) \cdot F^{-1} =$$
$$C_Z \cdot Z \cdot e_0^2 / R_{Strong} + Z \cdot (Z - 1) \cdot e_0^2 / R_{Strong} \qquad (7)$$

The first term on the right-hand side is the Coulomb repulsion for an incoming proton; the second term relates to the mutual Coulomb repulsion of Z protons in the nucleus. $U_0$ is the depth of the potential and is put equal to $A \cdot E_B \cdot R_{Strong}{}^2$; $C_Z$ is a proportionality factor and $E_B$ the binding energy per nucleon ($E_B = 8$ MeV, if $A \geq 12$ and $E_B < 8$ MeV, if $A < 12$). We are able to rescale $\omega_0$ such that $U_0$ vanishes. We put $A = 2 \cdot Z$ and multiply both sides with $(2 \cdot Z)^{1/3}$:

$$C_Z = e_0{}^2 \cdot 10^{13} / (1.2 \cdot \sqrt[3]{2}); \ \kappa = 1 + 2/3 \quad (8)$$

A least-squares fit of all available *LAD* yielded $\kappa = 1.659$ instead of $1 + 2/3$. This might be due to some crude assumptions: we have assumed that $M_{Proton} = M_{Neutron}$ and neglected the spin-orbit coupling. If the neutron number is slightly higher than Z ($A = 2 \cdot Z + \varepsilon_N$), then $C_Z$ contains a correction term (the spherical symmetry still holds):

$$E_{Th} = F \cdot C_Z \cdot Z^{\kappa} / (1 + \varepsilon_N / 2 \cdot Z) \quad (9)$$

### 2.2 Total nuclear cross-section $Q^{tot}$

With regard to the determination of $Q^{tot}$, we have borrowed elements from the collective model of nuclei and from the extended nuclear-shell theory (see Appendix A). First of all, we have to know three types $Q^{tot}{}_{type}$ of $Q^{tot}$: 1. $Q^{tot}{}_{max}$ required for the calculation of other quantities (Figure 1: $Q^{tot}{}_{max}$ amounts to 541 mb and $E_{res} = 20.12$ MeV), 2. $Q_c$ required for the calculation of the transition point from the Gaussian behavior to the exponential behavior of $Q^{tot}$, 3. $Q^{tot}{}_{as}$ determines the asymptotic behavior of Q (Figure 1: 299 mb). We start with 'Ansatz' (explanations will be given thereafter):

$$Q^{tot}(type) = a \cdot A + b \cdot A^{2/3} + c \cdot A^{1/3} + d \cdot Z^{\kappa} / A^{1/3} \quad (10)$$

**Table 3.** Parameters a - d of each type of $Q^{tot}{}_{(type)}$ of Eq. (10).

| $Q^{tot}{}_{(type)}$ | a | b | c | d |
|---|---|---|---|---|
| $Q^{tot}{}_{max}$ | 2.61696075942438 | 81.2923967886543 | 2.94220517608668 | - 1.95238820051575 |
| $Q^{tot}{}_c$ | 2.61323819764975 | 76.4164500007471 | 2.40550058121611 | - 1.26209790271275 |
| $Q^{tot}{}_{as}$ | 0.26244059384442 | 46.6811789688200 | 0.37714379933853 | - 0.14166405273391 |

Due to $R_{Strong} \sim A^{1/3}$ we obtain the following properties:

*Term a*

*This term incorporates a connection of $Q^{tot}_{(type)}$ to the volume of the nucleus. It is important in the resonance domain.*

*Term b*

*Proportionality to the area of the geometric cross-section: Potential scatter (major part), rotations by Coulomb repulsion/strong-interaction attraction (elastic) and nuclear reactions by changing the spin/isospin multiplicity (inelastic).*

*Term c*

*Proportionality to $R_{Strong}$: Collective excitations of the nucleus via spin-orbit-coupling, change of the angular momentum of the nucleus, inelastic resonance effects and elastic spin-spin scatter.*

*Term d*

*This term incorporates excitations of nuclear vibrations by Coulomb repulsion (resonance effect, inelastic) and elastic scatter.*

It should be pointed out that the asymptotic behavior of $Q^{tot} = Q^{tot}_{as}$ is mainly characterized by the term *b*, whereas the contribution of the other terms significantly decreased.

This connection mainly contains the term b above. We have verified its validity up to Z = 40.

According to results of extended nuclear-shell theory and *LAD*, $E_{res}$ is determined by:

$$\left.\begin{array}{l} E_{res} = E_{Th} + 11.94 + 0.29 \cdot (A-12) \ (A \geq 12) \\ E_{res} = E_{Th} + 11.94 \cdot [(A-1)/11]^{2/3} \ (A<12) \end{array}\right\} \text{(11)}$$

The shape of $Q^{tot}$ for media other than oxygen (Fig. 1) is more or less identical, and it can be stated by the formulas:

$$\left.\begin{array}{ll} Q^{tot} = Q^{tot}_{max} \cdot [\exp(-(E-E_{res})^2 / \sigma_{res}^2) - A_{Th}] \cdot [1 - A_{Th}] & (if \ E_{res} \geq E_{Th}) \\ Q^{tot} = Q^{tot}_{max} \cdot \exp(-(E-E_{res})^2 / 2\sigma_{res}^2) & (if \ E_{res} < E < E_c) \\ Q^{tot} = Q^{tot}_c - (Q^{tot}_c - Q^{tot}_{as}) \cdot \tanh((E-E_c)/\sigma_{as}) & (if \ E > E_c) \end{array}\right\} \text{(12)}$$

The parameters of Eq. (12) are given by:

$$\left. \begin{array}{l} \boldsymbol{\sigma_{res}} = (\boldsymbol{E_{res}} - \boldsymbol{E_{Th}}) \cdot \sqrt{\boldsymbol{\pi}} \, ; \; \boldsymbol{E_c} = \boldsymbol{E_{res}} + \sqrt{-2 \cdot \ln(\boldsymbol{I_c})} \\[4pt] \boldsymbol{\sigma_{as}} = \boldsymbol{\sigma_{res}} \cdot (\boldsymbol{Q^{tot}c} - \boldsymbol{Q^{tot}as}) / (\boldsymbol{Q^{tot}_{max}} \cdot \sqrt{-2 \cdot \ln(\boldsymbol{I_c})}) \\[4pt] \boldsymbol{A_{Th}} = \exp(-(\boldsymbol{E_{Th}} - \boldsymbol{E_{res}})^2 / \boldsymbol{\sigma_{res}}^2) \end{array} \right\} \quad (13)$$

$\sigma_{res}$ and $\sigma_{as}$ are determined by the continuity condition; for $E = E_c$ we have $Q^{tot} = Q^{tot}_c = Q_{max} \cdot I_c$ and by the identity of the first derivation at this position. $Q^{tot}_c$ or $I_c$ is obtained by the same method as given by Eq. (12). The analytical fit of Eq. (12) has to account for the continuity at $E = E_c$ and $Q^{tot} = Q^{tot}_c$ ($Q^{tot}_c = Q^{tot}_{max} \cdot I_c$) and the compatibility of the first derivative, i.e. $dQ^{tot}/dE$ at $E = E_c$.

### 2.3 Collective excitations based on the harmonic oscillator and the related potential problems

The starting point is the Schrödinger equation of the 3D harmonic oscillator:

$$\left. \begin{array}{l} \boldsymbol{H_{osc}} = \displaystyle\sum_{k=1}^{3} [\, \boldsymbol{p_k}^2 / 2M + \frac{M}{2} \cdot \boldsymbol{\omega_0}^2 \cdot \boldsymbol{q_k}^2 \,] \\[10pt] \boldsymbol{p_k} \Rightarrow \dfrac{\hbar}{i} \cdot \partial / \partial \boldsymbol{q_k} \end{array} \right\} \quad (14)$$

Since we mainly consider harmonic-oscillator eigen-functions, we only deal with the configuration-space representation, i.e., Hermite polynomials multiplied with a Gaussian, which form a complete set of functions:

$$\left. \begin{array}{l} \boldsymbol{\psi_{j,k,l}} = N \cdot \boldsymbol{H_j}(\boldsymbol{\rho_1}) \cdot \boldsymbol{H_k}(\boldsymbol{\rho_2}) \cdot \boldsymbol{H_l}(\boldsymbol{\rho_3}) \cdot e^{-\boldsymbol{\rho}^2 / 2} \\[4pt] \boldsymbol{\rho_k} = \boldsymbol{q_k} \cdot \sqrt{M \boldsymbol{\omega_0} / \hbar} \, ; \; \boldsymbol{\rho}^2 = \boldsymbol{\rho_1}^2 + \boldsymbol{\rho_2}^2 + \boldsymbol{\rho_2}^2 \\[4pt] \boldsymbol{E_{j,k,l}} = \hbar \boldsymbol{\omega_0} \cdot (3/2 + j + k + l) \;\; (j,k,l = 0,1,..,\infty) \end{array} \right\} (15)$$

These eigen-functions can be connected to the spherical harmonics (angular momentum) as well as the SU$_3$ classification (Elliott 1963, Meyer-Göppert and Jensen 1970). The above model is the physical base for various nuclear models. Thus, it can be used to describe rotations, vibrations, and excitations (with change of spin multiplicity) of nuclei. In particular, the angular momentum is conserved by Eq. (14) and the solution functions (15). The oscillator potential, however, would imply always stable nuclei and nuclear reactions with a change of the spin and isospin multiplicity could not occur. Even by extending Eq. (14) to a many-particle equation and to a Slater determinant (Hartree-Fock), the problem of nuclear reactions cannot be solved. Figure 3 shows this problem in the case of oxygen, where the abscissa is expressed in units of r = R$_{Strong.}$. The whole potential function V(r) can be expressed in terms of two different Gaussians (if the binding energy per nucleon $E_B$ amounts to ca. 7 – 8 MeV, which is true for most nuclei, the related potential V is nearly identical) :

$$\left.\begin{array}{l} V(r) = V_0 \cdot e^{-r^2/\sigma_0^2} + V_1 \cdot e^{-r^2/\sigma_1^2} \\ \sigma_0 = 0.43; \ \sigma_1 = 3.374; \ V_0 = -27.76 \, MeV; \ V_1 = 7.759 \, MeV \end{array}\right\} \ (16)$$

$$\left.\begin{array}{l} r_c \approx \frac{1}{\sqrt{2}} \cdot \sigma_0 \cdot \sigma_1 \cdot \sqrt{(-\sigma_1^2 \cdot V_0 - \sigma_0^2 \cdot V_1 \cdot A_\sigma)/(-\sigma_1^4 \cdot V_0 - \sigma_0^4 \cdot V_1 \cdot A_\sigma)} \\ r_c \approx 0.4582; \qquad A_\sigma = \exp[(\sigma_1^2 - \sigma_0^2)/2 \cdot \sigma_1^2)] \end{array}\right\} \ (17)$$

Only for $r \le r_c = 0.4582$ (this is the domain with positive curvature of the composite Gaussians according to Eq. (16)) is a harmonic-oscillator approach useful, i.e., a Taylor expansion of Eq. (16) provides:

$$V(r) \approx V_0(1 - r^2/\sigma_0^2) + V_1(1 - r^2/\sigma_1^2) \quad (r \le r_c) \quad (18)$$

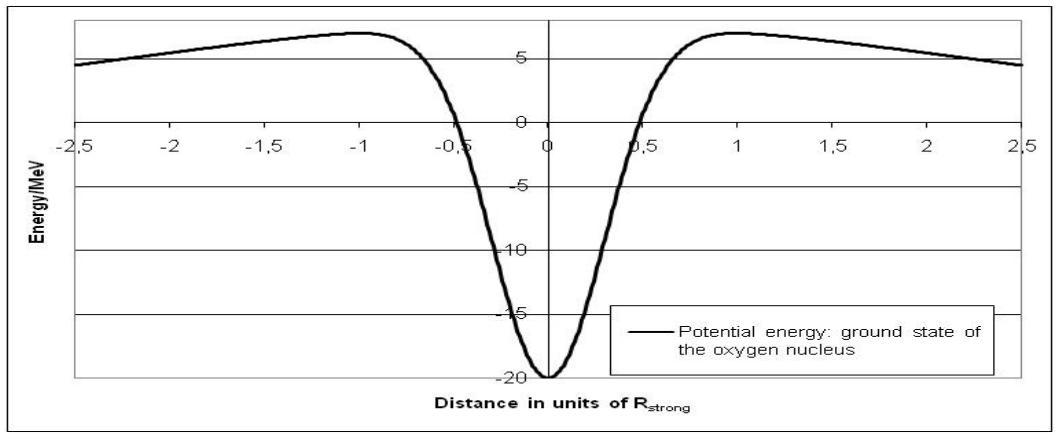

**Figure 3.** Effective potential of the oxygen nucleus

For $r_c < r < 1$, strong interactions are present with decreasing tendency, whereas Coulomb repulsion increases (if $r \ge 1$, strong interactions are negligible). According to Feynman and Schwinger (Feynman 1962), strong interactions can be described by a Gaussian potential rather than a Yukawa potential, which leads to singularities at $r \to 0$. The 3D oscillator concept is only partially adequate for the treatment of many-particle problems, which represent nonlinear interacting systems, in nuclear physics. Thus, only the ground state and low excited states are appropriately calculated by harmonic oscillator wave-functions with perturbation methods for the deviations of oscillator potential from a Gaussian. High excited states and, in particular, nuclear reactions cannot be treated with perturbation methods based on harmonic oscillators. These aspects are given in Appendix A; for further details we refer to Ulmer et al 2009.

### 3. Results and conclusions

Based on results obtained by methods in Appendix A we have calculated parameters necessary to evaluate formulas (8 – 11) related to properties of the collective model and applied to nuclei starting with deuteron up to Z = 30 (zinc). A purpose of these theoretical calculations was, besides the foundation of the parameters of the collective model, a comparison with *LAD* (if possible) and a toolkit of cross-sections for Monte-Carlo calculations. A main result is that some of the 'secondary protons' are in reality primary protons, which have been scattered at the nuclear potential (elastic, if the nucleus remains unchanged or inelastic, if rotations, vibrations, or excited states are produced). The reaction protons, which release neutrons, additional protons, and other particles via clusters amount to 1 % (for $E_0 \approx 100$ MeV) and to 4 % (for $E_0 \approx 250$ MeV) of the impinging proton beam. Under certain restrictions, potential/resonance scatter can be described by the Breit-Wigner-Flügge formula (Segrè 1964). The resonances play a decisive role with regard to $E_{res}$ and the cross-section in this environment. A particular impact is the scatter of protons in media like bone, implants, and collimators (Figure 4).

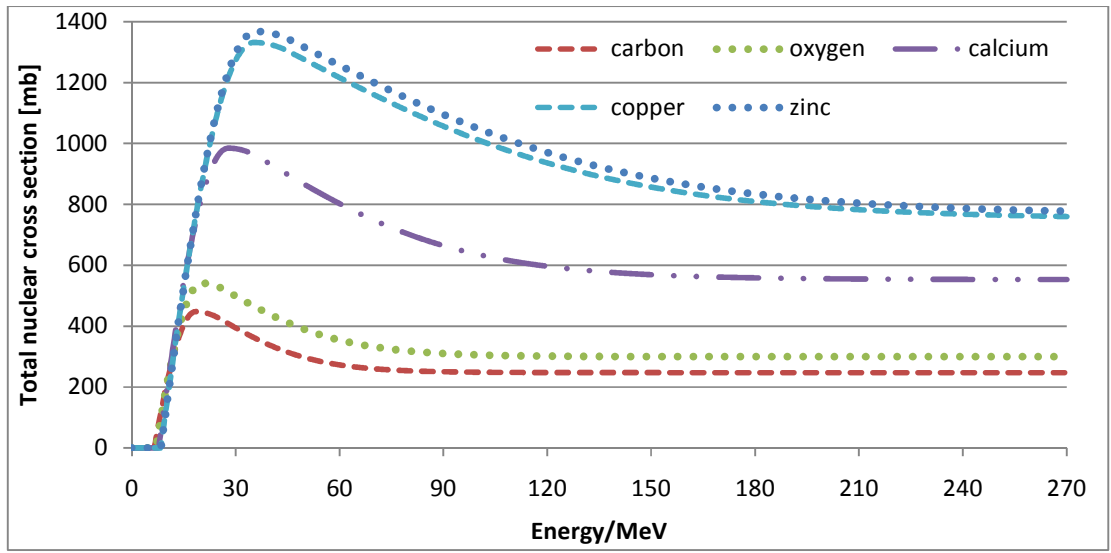

**Figure 4.** Total nuclear cross-section for some nuclei (C, O, Ca, Cu, Zn).

Since water serves as a reference material in radiotherapy, we have listed the most important nuclear reactions and secondary protons emerging from the proton – oxygen interaction:

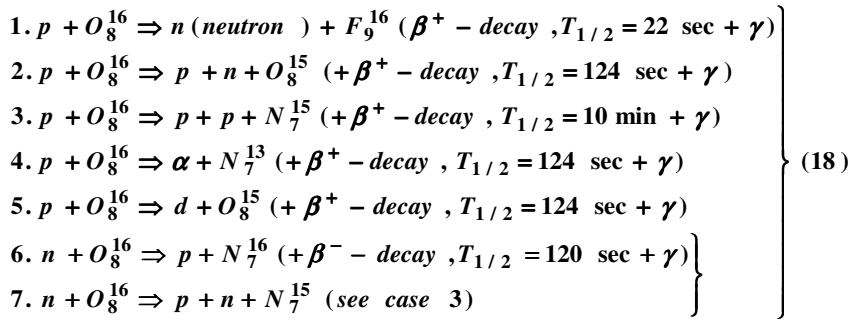

$$
\left.
\begin{array}{l}
\textbf{1.}\ p + O_8^{16} \Rightarrow n\ (neutron\ ) + F_9^{16}\ (\boldsymbol{\beta^+ - decay}\ , T_{1/2} = 22\ \sec + \boldsymbol{\gamma}) \\[4pt]
\textbf{2.}\ p + O_8^{16} \Rightarrow p + n + O_8^{15}\ (+\boldsymbol{\beta^+ - decay}\ , T_{1/2} = 124\ \sec + \boldsymbol{\gamma}\ ) \\[4pt]
\textbf{3.}\ p + O_8^{16} \Rightarrow p + p + N_7^{15}\ (+\boldsymbol{\beta^+ - decay}\ , T_{1/2} = 10\ \min + \boldsymbol{\gamma}) \\[4pt]
\textbf{4.}\ p + O_8^{16} \Rightarrow \alpha + N_7^{13}\ (+\boldsymbol{\beta^+ - decay}\ , T_{1/2} = 124\ \sec + \boldsymbol{\gamma}) \\[4pt]
\textbf{5.}\ p + O_8^{16} \Rightarrow d + O_8^{15}\ (+\boldsymbol{\beta^+ - decay}\ , T_{1/2} = 124\ \sec + \boldsymbol{\gamma}) \\[4pt]
\textbf{6.}\ n + O_8^{16} \Rightarrow p + N_7^{16}\ (+\boldsymbol{\beta^- - decay}\ , T_{1/2} = 120\ \sec + \boldsymbol{\gamma}) \\[4pt]
\textbf{7.}\ n + O_8^{16} \Rightarrow p + n + N_7^{15}\ (see\ case\ \ 3)
\end{array}
\right\} \quad (18)
$$

All types of β+-decay emit one γ-quant; its energy is of the order 0.6 MeV – 1 MeV. The β+-decay of $F_9^{16}$ has a half-life of about 20 seconds, and further γ-quanta are produced by collisions of positrons with environmental electrons. The remaining heavy recoil fragments have partially half-times up to 10 minutes ($N^{15}$). Since Figures 1 and 4 refer to the total nuclear cross-section in dependence of the actual (residual) proton energy, we have to add some qualitative aspects on the 5 different types with regard to the required proton energy: If E < 50 MeV the type (1) is the most probable case with rapid decreasing tendency between 50 MeV < E < 60 MeV to become zero for E > 60 MeV. Type (2) also pushes out a neutron, but the incoming proton is not absorbed; the required energy amounts, at least, to 50 MeV. Type (3) is similar, but requires, at least, ca. 60 MeV with increasing probability. The release of α-particles resulting from clusters in the nucleus requires E ≈ 100 MeV and the probability is increasing up to E ≈ 190 MeV; thereafter it is decreasing rapidly, since higher energy protons destroy these clusters by pushing out deuterons according to type 5. Thus case 5 is energetically possible for E > 60 MeV, but the significance is only increasing for E > 200 MeV. The nuclear reactions 6 and 7 result from the release of neutrons; they may undergo further interactions with oxygen. According to Ulmer et al (2009) the cases 1 and 6 are noteworthy. Thus the neutron released by the incoming proton has not absolutely to be the result of a real collision (threshold energy, at least, 20 MeV). For energies E > 7 MeV and E < 20 MeV a resonance effect via exchange interaction between proton and nucleus via a $\pi^-$ meson (Pauli principle) is also possible. By that, the incoming proton leaves the oxygen nucleus as a neutron. Case 6 represents the reversal process, i.e., the incoming (secondary) neutron is converted to a proton via $\pi^+$ exchange. *The calculation procedure of the stopping power $S_{sp,r}$ of reaction protons is presented in the Appendix B* (see also Ulmer et al 2009).

We should also point out that nuclear reactions stated in the listing (18) are only a part of the total decrease of the primary proton fluence according to Eq. (2). If E < 100 MeV, the main part of the decrease of the primary proton fluence results from nuclear scatter of protons by the oxygen nucleus (deflection of primary protons without release of further nucleons) via intermediate deformations and oscillations of the nucleus. These oscillations are damped by emission of γ-quanta with very low energy (ca. 1 keV), which are most widely absorbed by the Auger effect. The most important source for recoil protons are elastic collisions of projectile protons with the proton of Hydrogen. Released neutrons of type 1 also loose most widely their energy by such collisions before they become thermal neutrons. These neutrons usually escape and undergo β- - decay to produce a proton, electron and a γ-quant (0.77 MeV), $T_{1/2}$ = 17 min. Neutrons of type 2 carry a much higher energy and can escape without any significant collisions. The cases 6 and 7 referring to the neutron interaction with the oxygen nucleus are the only noteworthy inelastic contributions. Figure 1 shows the dose contribution of reaction protons for some therapeutic proton energies. The contributions of deuterons and α-particles are also accounted for in this figure. The tails beyond the ranges $R_{CSDA}$ mainly result from the reaction types 6 and 7 of the listing (18).

Figure 5 is the correspondance to Figure 2 with particular importance to the passage of protons to bone and collimator scatter. The mean standard deviations of the calculated results according to Figure 4 from LAD according to Chadwick et al (1996) amount to 0.63 % (carbon and oxygen), 0.73 % (calcium), 0.71 % (copper), and 0.72 % (zinc).

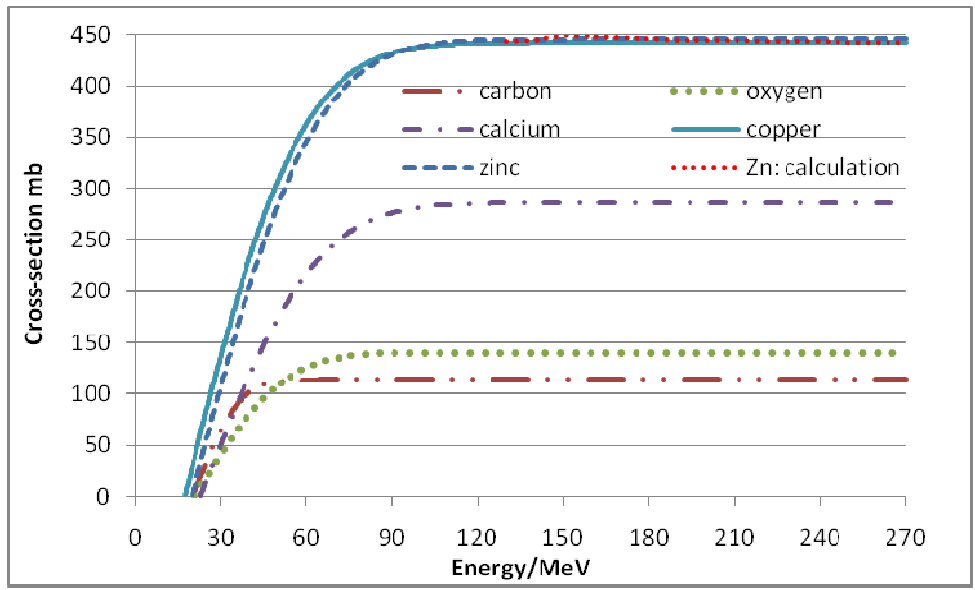

**Figure 5.** Nuclear reaction part of the cross-sections of the nuclei C, O, Ca and Zn (smooth curves).

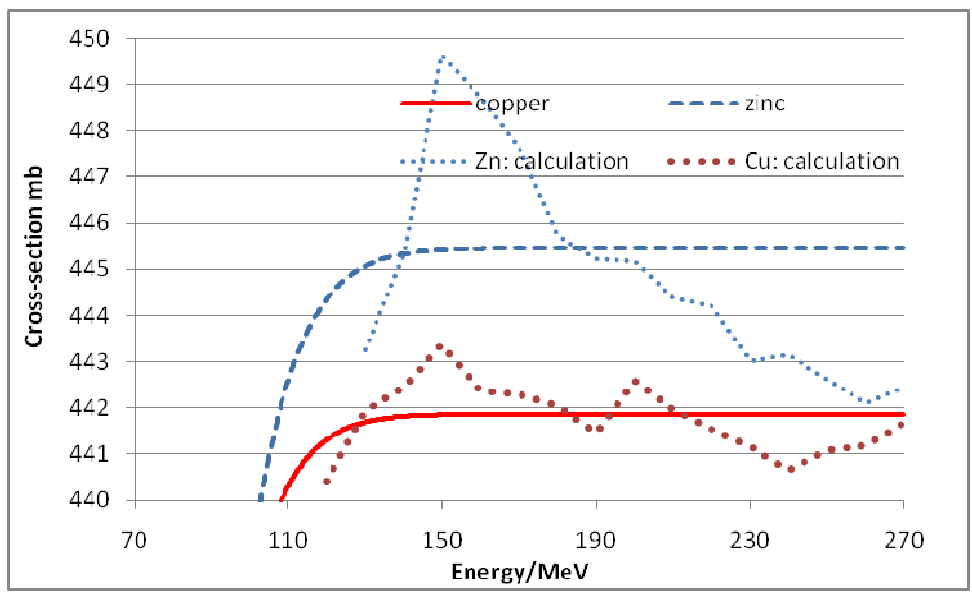

**Figure 6.** Section of Figure 5: Calculated contributions (extended nuclear shell theory) of and smooth curves.

Figures 5 and 6 represent the nuclear reaction contributions $Q^{tot}_r$ of the total nuclear cross-sections $Q^{tot}$. The mean standard deviations amount to 0.3 % (carbon), to 0.32 % (oxygen), to 0.27 % (copper), and to 0.31 % (zinc). However, the deviations of the smoothed curves are not significant with regard to the overall behavior of the protons. The fitting curves are obtained by the following formula:

$$\left. \begin{array}{l} Q^{tot}{}_r = Q^{tot}{}_{as} \cdot erf \left( \left( Energy - E_{pot} \right) / \sigma_r \right) \cdot f_c \\ f_c = a \cdot A^p + b \cdot A + 0.438 \\ a = 0.00148215 \qquad p = 1.07754 \qquad b = 1.02 \cdot 10^{-9} \end{array} \right\} \quad (19)$$

The parameters $\sigma_r$ and $E_{pot}$ cannot be determined by a simple formula, since $E_{pot}$ refers to the potential depth for neutrons and $\sigma_r$ to a *rms* value of the energy distribution of nuclear reactions (the cases under discussion are presented by Table 4). A crude approximation of the function $f_c$ would be 4/9 independent of the nuclear mass number A. Since $E_{pot}$ depends on the isotope and can rather be different for certain isotopes (e.g. if the number of protons and neutrons is even), we are obliged to use the statistical weight of the isotopes to determine the effective $E_{pot}$ according to formula (19).

**Table 4.** Parameters of Eq. (19)

|  | C | O | Ca | Cu | Zn |
|---|---|---|---|---|---|
| $E_{pot}$ in MeV | 20.41 | 20.97 | 24.99 | 17.35 | 19.79 |
| $\sigma_r$ in MeV | 27.17 | 34.14 | 44.78 | 44.92 | 46.81 |

Figure 7 clearly shows that the passage of protons through other media than water (reference medium in therapy planning) cannot only handled by a path-length correction, but the fluence decrease has also to be accounted for.

The transport of secondary reaction protons resulting from the spectral distribution of these protons has to be taken into account; and the overall spectral distributions rather obey a Landau than a Gaussian distribution (Figures 8 and 9). The tails at $z \geq R_{CSDA}$ result from tertiary protons induced by neutrons and the resonance interaction via meson exchange as pointed out in a previous section. Some further implications of the secondary protons are buildup effects of Bragg curves. This behavior of reaction protons along the proton track represents a principal question in understanding the physical foundation of Bragg curves. The calculation procedure to Figures 8 – 9 is presented in Appendix B.

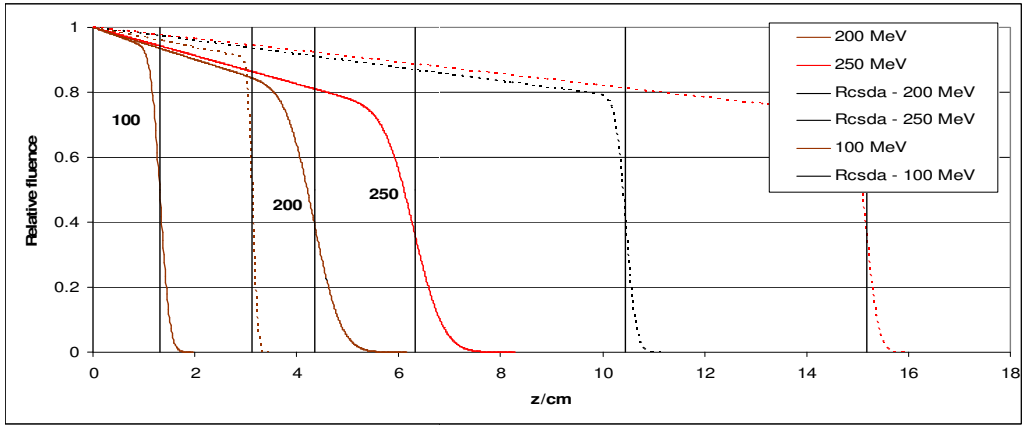

**Figure 7.** Fluence decrease of Ca (dots) und Cu (solids) from Figure 4

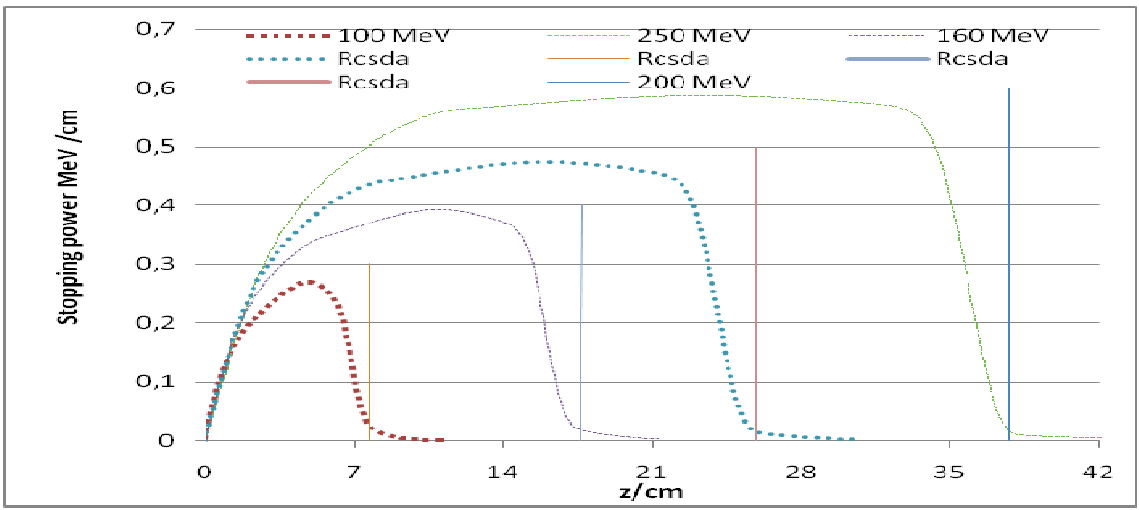

**Figure 8.** Stopping power (in water) of secondary/tertiary protons for the initial proton energies 100 MeV, 160 MeV, 200 MeV and 250 MeV. The $R_{CSDA}$ ranges are indicated by perpendicular straight lines.

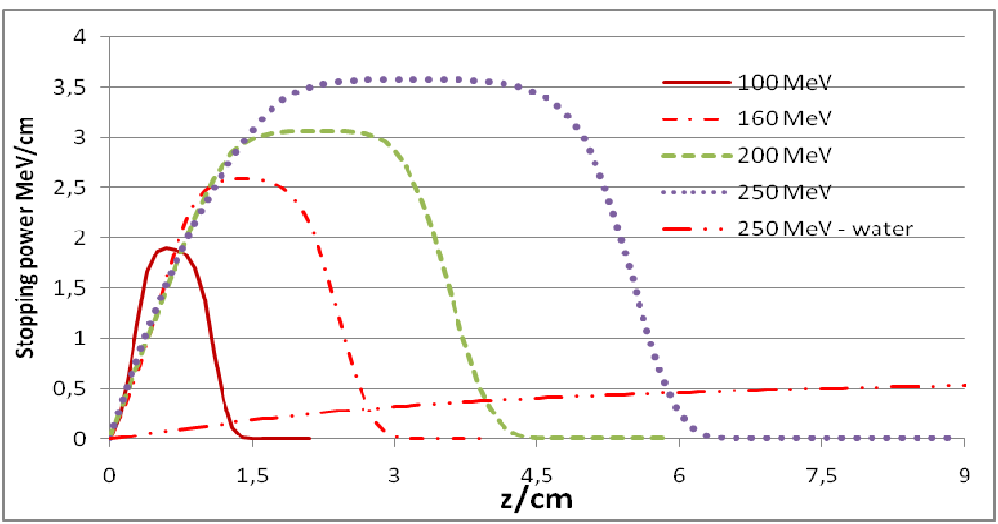

**Figure 9.** Stopping power (in copper and comparison with water) of secondary/tertiary protons for the initial proton

energies 100 MeV, 160 MeV, 200 MeV and 250 MeV.

**Appendix A: Aspects of the extended nuclear-shell theory (online part)**

Let us first consider the usual Schrödinger equation for a bound system:

$$\left. \begin{array}{l} E \cdot \psi + \dfrac{\hbar^2}{2 \cdot M} \cdot \Delta \psi = \varphi(x,y,z) \cdot \psi \\[2mm] \varphi \leq 0 \ (x,y,z \in R_3); \ \Delta : 3D - Laplace \ operator \end{array} \right\} \quad (20)$$

A nonlinear Schrödinger equation is obtained by introducing the potential φ, proportional to the density of solutions:

$$\varphi = \lambda \cdot \left| \psi(x,y,z) \right|^2 = \lambda \cdot \int \delta(x-x') \cdot \delta(y-y') \cdot \delta(x-x') \cdot \left| \psi(x',y',z') \right|^2 dx' dy' dz' \quad (21)$$

During the past decades, this type has been encountered in many fields of physics, such as superconductivity, nuclear and plasma physics (see Ulmer et al 2009, Milner 1990, and various other references). The coupling constant λ is negative (in which case, the solutions are bound states with E < 0); Eq. (21) can be interpreted as a contact interaction. It is known from many-particle problems (e.g., quantum electrodynamics, Hartree/Hartree-Fock method, etc) that the mutual interactions between the particles in configuration space lead to nonlinear equations in quantum mechanics. However, in these cases, there are not at all contact interactions; the nonlinear Schrödinger equation above is an idealistic case. By taking ε → 0, the Gaussian kernel is transformed into a δ kernel:

$$\varphi = \lambda \cdot \int (1/(\sqrt{\pi}^3 \cdot \varepsilon^3) \cdot \exp[-((x-x')^2 + (y-y')^2 + (z-z')^2)/\varepsilon^2] \left| \psi(x',y',z') \right|^2 dx' dy' dz' \quad (22)$$

The nonlinear/nonlocal Schrödinger equation can be interpreted as a self-interaction of a many-particle system with internal structure, and it is possible to generalize this type by incorporation of additional internal symmetries (e.g., the introduction of the spin to obtain spin-orbit coupling, $SU_2$, $SU_3$, and also discrete-point groups). According to the principles developed in Ulmer et al 2003 and 2009, we are able to write Eq. (22) in the form of an operator equation (the Gaussian kernel is Green's function):

$$\varphi = \lambda \cdot \exp(\frac{1}{4} \cdot \varepsilon^2 \cdot \Delta) \cdot \left| \psi(x,y,z) \right|^2 \quad (23)$$

Expanding this operator in terms of a Lie series and keeping only the terms up to Δ, Eq. (24) becomes a stationary Klein-Gordon equation, which describes the interaction between the particles obeying the Ψ-field:

$$\left. \begin{array}{l} \exp(\frac{1}{4} \cdot \varepsilon^2 \cdot \Delta) = 1 + \frac{1}{4} \cdot \varepsilon^2 \cdot \Delta + 0(higher-order\ terms) \\ (1 + \frac{1}{4} \cdot \varepsilon^2 \cdot \Delta) \cdot \varphi = \lambda \cdot |\psi|^2 \end{array} \right\} \quad (24)$$

By rescaling the Klein-Gordon equation, we obtain the more familiar form: $1 + 0.25\ \varepsilon^2\ \Delta \rightarrow k^2 + \Delta$; Green's function is of the form:

$$\left. \begin{array}{l} G(r,r') = N \cdot \exp(-k \cdot |r - r'|) \cdot \dfrac{1}{|r-r'|} \\[2mm] N : normalization\ factor \end{array} \right\} \quad (25)$$

By setting k $\rightarrow$ 0, the Poisson equation of electrostatics is obtained, if $|\psi|^2$ is interpreted as a charge density. The Gaussian kernel K also represents the exchange of virtual particles between the nucleons. In view of this fact, we point out that we have incorporated a many-particle system from the beginning. Which information now does this nonlinear/nonlocal Schrödinger equation provide? In order to obtain a connection of the combined equations (20 22) with the oscillator model of nuclear shell theory, we analyze the kernel K in detail. In the Feynman-propagator method (see Feynman 1962 and references therein), the expansion of K in terms of generating functions is an important tool:

$$\left. \begin{array}{l} K(\varepsilon, x'-x, y'-y, z'-z) = (1/\sqrt{\pi}^3 \cdot \varepsilon^3) \displaystyle\sum_{n1=0}^{\infty} \exp(-x'^2/\varepsilon^2) \cdot H_{n1}(x'/\varepsilon) \cdot x^{n1}/(\varepsilon^{n1} \cdot n1!) \cdot \\[3mm] \cdot \displaystyle\sum_{n2=0}^{\infty} \exp(-y'^2/\varepsilon^2) \cdot H_{n2}(y'/\varepsilon) \cdot y^{n2}/(\varepsilon^{n2} \cdot n2!) \cdot \\[3mm] \cdot \displaystyle\sum_{n3=0}^{\infty} \exp(-z'^2/\varepsilon^2) \cdot H_{n3}(z'/\varepsilon) \cdot z^{n3}/(\varepsilon^{n3} \cdot n3!) \end{array} \right\} \quad (26)$$

Inserting this expression into the nonlinear/nonlocal Schrödinger equation, we obtain:

$$\left. \begin{array}{l} E \cdot \psi + \dfrac{\hbar^2}{2 \cdot M} \cdot \Delta \psi = \varphi(x,y,z) \cdot \psi = \lambda \cdot (1/\sqrt{\pi}^3 \cdot \varepsilon^3) \displaystyle\sum_{n1=0}^{\infty} \sum_{n2=0}^{\infty} \sum_{n3=0}^{\infty} \Phi_{n1,n2,n3} \cdot x^{n1} \cdot y^{n2} \cdot z^{n3} \cdot \psi \\[3mm] \Phi_{n1,n2,n3} = \dfrac{1}{n1!} \cdot \dfrac{1}{n2!} \cdot \dfrac{1}{n3!} \cdot \dfrac{1}{\varepsilon^{n1+n2+n3}} \cdot \int |\psi(x',y',z')|^2 \cdot \exp(-(x'^2+y'^2+z'^2)/\varepsilon^2) \cdot \\[3mm] \cdot H_{n1}(x'/\varepsilon) \cdot H_{n2}(y'/\varepsilon) \cdot H_{n3}(z'/\varepsilon) dx' dy' dz' \end{array} \right\} \quad (27)$$

The equation above represents a highly anharmonic oscillator equation of a self-interacting field. Since the square of the wave-function is always positive definite, all terms with odd numbers of n1, n2, and n3 vanish due to the anti-symmetric properties of those Hermite polynomials. For $r_c \leq \varepsilon/\sqrt{2}$ (domain with positive curvature), the whole equation is reduced to a harmonic oscillator with self-interaction; the higher-order terms are small perturbations. We summarize the results and refer to previous publications,

(Ulmer et al 1978, Ulmer 1980, and Milner 1990):

$$E \cdot \psi + \frac{\hbar^2}{2 \cdot M} \cdot \Delta \psi = \varphi(x,y,z) \cdot \psi = \lambda \cdot (1/\sqrt{\pi}^3 \cdot \varepsilon^3) \cdot [\Phi_{0,0,2}(x^2 + y^2 + z^2) + \Phi_{0,0,0}] \cdot \psi \quad (28)$$

The solutions of this equation are those of a 3D harmonic oscillator; the classification of the states by $SU_3$ and all previously developed statements with regard to the angular momentum are still valid. The only difference is that the energy levels are not equidistant; this property can easily be verified. The usual ground state energy is given by $3 \cdot \hbar \omega_0 / 2$. This energy level is lowered by the term $\sim \lambda \cdot \Phi_{0,0,0}$, depending on the ground-state wave-function. The energy difference between the ground and the first excited state amounts to $\hbar \omega_0$; this is not true in the case above, since the energy level of the excited states depends on the corresponding eigen-function (these are still the oscillator eigen-functions!). Next, we will include the terms of the next order, which are of the form $\sim \lambda \cdot (\Phi_{0,2,2}, \Phi_{2,2,0}, \Phi_{2,0,2})$:

$$\left. \begin{array}{l} E \cdot \psi + \frac{\hbar^2}{2 \cdot M} \cdot \Delta \psi = \varphi(x,y,z) \cdot \psi = \lambda \cdot (1/\sqrt{\pi}^3 \cdot \varepsilon^3) \cdot [\Phi_{0,0,0} + \Phi_{0,0,2}(x^2 + y^2 + z^2) + T] \cdot \psi \\ T = \Phi_{2,2,0} \cdot x^2 \cdot y^2 + \Phi_{2,0,2} \cdot x^2 \cdot z^2 + \Phi_{0,2,2} \cdot y^2 \cdot z^2 \end{array} \right\} \quad (29)$$

The additional term T represents tensor forces. The whole problem is still exact soluble. In further extensions of the nonlinear/nonlocal Schrödinger equation, we are able to account for spin, isospin, and spin-orbit coupling. The spin-orbit coupling, as an effect of an internal field with nonlocal self-interaction, is plausible, since the extended nucleonic particle has internal structure; consequently, we have to add $H_{so}$ to the nonlinear term ($g_\tau$ represents the magnetic moment of the proton/neutron as a coupling constant):

$$H_{so} \cdot \psi = g_\tau \cdot \frac{\hbar \cdot \sigma}{4 \cdot M \cdot c^2} \cdot \nabla \varphi \cdot x \, p \cdot \psi \quad (30)$$

$\Psi$ is now (at least) a Pauli spinor (i.e., a two-component wave-function), and together with $H_{so}$ the $SU_3$ symmetry is broken. We should like to point out that the operation $\nabla \varphi$ acts on the Gaussian kernel K:

$$\nabla \varphi = -\frac{2}{\varepsilon^2} \cdot [H_1((x-x')/\varepsilon), H_1((y-y')/\varepsilon), H_1((z-z')/\varepsilon)] \cdot \varphi \quad (31)$$

The expression in the bracket of the previous equation represents a vector, and p ($p \rightarrow -i \cdot \hbar \cdot \nabla$) acts on the wave-function. Since the neutron is not a charged particle, the spin-orbit coupling of a neutron can only involve the angular momentum of a proton. In nuclear physics, these nonlinear fields are adequate for the analysis of clusters (deuteron, He, etc.). Milner (1990) has extended the theory to describe nuclei with odd spin.

The complete wave-function $\Psi_c$ is now given by the product of a function in configuration space $\Psi$ multiplied with the total spin and isospin functions.

We should like to add that an extended harmonic oscillator model with tensor forces has been regarded in Elliott (1963). The application of oscillator models in nuclear physics goes back to Heisenberg (1935); Feynman and Schwinger, (see Feynman 1962), have verified that the use of Gaussians in the description of meson fields provides many advantages compared to the Yukawa potential (Green's function according to Eq. (25)). In a final step, we consider the generalized Hartree-Fock method (HF) to solve the many-particle problem. In order to derive all required formulas, it would be convenient to use second quantization of the nonlinear/nonlocal Schrödinger equation. The nonlinear/nonlocal Schrödinger equation with Gaussian kernel for the description of the strong interaction, including the spin-orbit coupling, can be written by Fermion field operators, leading from an extended particle with internal structure to a many-particle theory. Thus the method of second quantization is only suitable to derive the calculation procedure: extension of the Pauli principle to isospin besides spin, inclusion of spin-orbit coupling, and exchange interactions. This is the consequence of dealing with identical particles, in which case every state can only occupy one quantum number. In order to get numerical results (i.e., the minimum of the total energy of an ensemble of nucleons, the extraction of the excited states, the scatter amplitudes, etc.), we have to use representations of the wave-function by at least one determinant in the configuration space. Therefore we are able to avoid the 'language' of second quantization of fermions. Before we start to explain the calculations by including one or more configurations, we recall that, according to Figure 3, we have an increasing contribution of the Coulomb repulsion for $r > r_c$, though in the domain $r < r_c$, the contributions of the Coulomb interactions are negligible. Since all basis elements of the calculation procedures, i.e., the calculation of eigen-functions in the configuration space, two-point kernels of strong interactions between nucleons, and the spin-orbit coupling can be expressed in terms of Gaussians and Hermite polynomials, we want to proceed in the same fashion with regard to the Coulomb part. According to results of elementary-particle models (e.g., see Feynman 1972), the charge of the proton is located in an extremely small sphere with radius $r_p = 10^{-14}$ cm, not at one 'point'. Therefore, we write the decrease of the proton Coulomb potential by $1/(r+r_p)$; for $r = 0$, we then obtain $10^{14}$ cm$^{-1}$, not infinite. In a sufficiently small distance of $r = 2.4 \cdot 10^{-13}$ cm, we can approximate the Coulomb potential with high precision by:

$$\frac{1}{r+r_p} = c_0 \cdot \exp(-r^2/r_0^2) + c_1 \cdot \exp(-r^2/r_1^2) + c_2 \cdot \exp(-r^2/r_2^2) \quad (32)$$

The mean standard deviation amounts to $10^{-5}$, if the parameters of Formula (32) are chosen as:

$$\left.\begin{array}{l} c_0 = 0.5146 \cdot 10^{14}; \ c_1 = 0.3910 \cdot 10^{14}; \ c_2 = 0.0944 \cdot 10^{14} \\ r_0 = 0.392 \cdot 10^{-13} \, cm; \ r_1 = 0.478 \cdot 10^{-13} \, cm; \ r_2 = 2.5901 \cdot 10^{-13} \, cm \end{array}\right\} \ (33)$$

If necessary, it is possible to rescale $r_0$, $r_1$, and $r_2$ by dividing by $(A)^{1/3}$. The contribution with $c_2$ incorporates a long-range correction.

In the absence of an external electromagnetic field, the Hamiltonian reads as:

$$\left.\begin{array}{l} H = \sum_j - \frac{\hbar^2}{2M} \Delta_j + H_{so} + H_{Coul} + H_{strong} \\[2mm] H_{Coul} = e_0^2 \cdot \sum_{j, proton} \sum_{l, proton \neq j} \sum_{k=0}^{2} c_k \cdot \exp(-(r_j - r_l)^2 / r_k^2) \\[2mm] H_{strong} = -g_s \cdot \sum_j \sum_{l \neq j} \exp(-(r_j - r_l)^2 / \sigma_s^2) \end{array}\right\} \ (34)$$

Note that it is possible to distinguish between the proton and the neutron masses by indexing M; the $\varepsilon$, previously used in Eq. (26), has been replaced by $\sigma_s$. The coupling constant of $g_s$ is 1, if the Coulomb interaction is scaled to

$$\left.\begin{array}{l} g_s = 1 \\ e_0^2 / (\hbar \cdot c) = 1/137 \end{array}\right\} \ (35)$$

Thus, in theoretical units with $e_0 = c = h/2\pi = 1$, the coupling constant $g_s$ assumes 137. This relation can be best seen in the Dirac equation containing a Coulomb repulsion potential $\sim e_0^2$ and a strong interaction term $\sim -g_s$. The aforementioned relation is obtained by dividing the kinetic-energy operator $c \cdot \boldsymbol{\alpha} \cdot \boldsymbol{p} \ \rightarrow \ - \cdot i c \cdot \boldsymbol{\alpha} \cdot \boldsymbol{h} \cdot \nabla$ and $\boldsymbol{\beta} \cdot \boldsymbol{m} c^2$ by $(c \cdot \boldsymbol{h})$. In the calculations for deuteron, He[3], and He, we have assumed the range length $\sigma_s$:

$$\left.\begin{array}{l} \sigma_s = \sigma_{sp} = \frac{\hbar}{m_p \cdot c} \approx 10^{-13}\ cm \\[2mm] m_p:\ mass\ of\ \pi - meson \end{array}\right\}\ (36)$$

This assumption turned out to be not sufficient; a replacement of $\sigma_s$ was justified to distinguish between the range length $\sigma_{sp}$ ($\pi$-mesons) and $\sigma_{sk}$ (K-mesons):

$$\left.\begin{array}{l} -g_S \cdot \exp(-(r_j - r_l)^2 / \sigma_s{}^2) \Rightarrow -g_S \cdot [c_{sp} \cdot \exp(-(r_j - r_l)^2 / \sigma_{sp}{}^2) + c_{sk} \cdot \exp(-(r_j - r_l)^2 / \sigma_{sk}{}^2)] \\[2mm] c_{sp} = 1 - (\sigma_{sk} / \sigma_{sp})^2;\ c_{sk} = (\sigma_{sk} / \sigma_{sp})^2 \\[2mm] \sigma_{sp} = 1.02 \cdot 10^{-13} cm;\ \sigma_{sk} = 0.29 \cdot 10^{-13}\ cm \end{array}\right\}\ (37)$$

The range length $\sigma_{sk}$ is proportional to $1/m_k$ ($m_k$: mass of the K-meson).

The HF method provides the best one-particle approximation of the closed-shell case.

$$\Phi = \frac{1}{\sqrt{N!}} \begin{vmatrix} \varphi_{k1}(1)\ldots\ldots & \ldots \varphi_{k1}(N) \\ \varphi_{k2}(1)\ldots\ldots & \ldots \varphi_{k2}(N) \\ \ldots\ldots\ldots & \ldots\ldots\ldots & \ldots\ldots \\ \varphi_{kN}(1)\ldots\ldots & \ldots \varphi_{kN}(N) \end{vmatrix}\ (38)$$

The one-particle functions $\varphi_{k1}(1)$, …, $\varphi_{kN}(N)$ contain all variables (configuration space of position coordinates, spin, and isospin). By using a complete system of trial functions, e.g., a Gaussian multiplied with Hermite polynomials, the HF limit is obtained. In view of our question to calculate the *S-matrix* and the cross section of the proton-nucleon interactions (elastic, inelastic, resonance scatter, and nuclear reactions), this restriction is insufficient. In particular, we have to add excited configurations and virtually-excited configurations. The role of excited states is clear. As an example, we regard the O nucleus, where the total spin is 0. If a proton or neutron of the highest-occupied shell is excited, then the spin may change, and both, highest-occupied and lowest-unoccupied shell, are occupied by one nucleon. The emitted nucleon may be regarded as a 'hole'. This procedure can be repeated to higher-unoccupied states and to linear combinations of configurations with different nucleon numbers. A virtually-excited state is produced, if the configuration of the excited state only formally exists for the calculation procedure, but cannot be reached physically. An example of this case is already the deuteron with isospin 0 and spin 1. An excited state with spin 1 or 0, where proton and neutron occupy different energy levels (shells), does not exist. In spite of this situation, the HF method does not provide the correct ground state, and linear combinations of determinants with different spin states (S = 1, -1, 0) and 'holes' have to be included. These virtual states also enter the calculation of the S-matrix and of the cross-section.

We have performed HF-configuration-interaction calculations (HF - CI) for the nuclei: deuteron, He[3], He, Be, C, Si, O, Al, Cu, and Zn. The set of basic functions comprises $2 \cdot (A + 13)$ functions with the following properties:

$$\varphi(x) = \sum_{j=0}^{N} [A_j \cdot H_j(\alpha_1 \cdot x) \cdot \exp(-\tfrac{1}{2} \cdot \alpha_1^2 \cdot x^2) + B_j \cdot H_j(\alpha_2 \cdot x) \cdot \exp(-\tfrac{1}{2} \cdot \alpha_2^2 \cdot x^2)]$$

$$\varphi(x,y,z) = \varphi(x) \cdot \varphi(y) \cdot \varphi(z)$$
$$N = A + 13$$

$$(39)$$

Both $\alpha_1$-functions and $\alpha_2$-functions are chosen such that the number of functions is A + 13. The different range parameters $\alpha_1$ and $\alpha_2$ are useful, since different ranges can be accounted for. If $\alpha \gg \beta$, the related wave-functions decrease much more rapidly (central part of the nucleus), whereas the $\beta$-contributions preferably describe the behavior in the domain $r \geq r_c$. With the help of this set of trial functions[1] (Ritz's variation principle), we obtain the best approximation of the total energy E by $E_{app}$ and the nuclear shell energies (for occupied and unoccupied shells). For bound states, $E_{app} > E$ is always fulfilled. It should be noted that for computational reasons it is useful to replace the set of functions (38) by the non-orthogonal set:

$$\varphi_{n1,n2,n3} = x^{n1} \cdot y^{n2} \cdot z^{n3} \cdot [A_{n1,n2,n3}(\alpha_1) \cdot \exp(-\tfrac{1}{2}\alpha_1^2 \cdot r^2) +$$
$$+ B_{n1,n2,n3}(\alpha_2) \cdot \exp(-\tfrac{1}{2}\alpha_2^2 \cdot r^2)] \qquad (40)$$

By forming arbitrary linear combinations depending on $\alpha_1$ and $\alpha_2$ we obtain the same results as by the expansion (38). The exploding coefficients of the Hermite polynomials are an obstacle in numerical calculations and can be avoided by the expansion (39). The minimal basis set for the calculation of deuteron would be one single trial function, i.e. a Gaussian without further polynomials. This is, however, a crude approximation and already far from the HF limit. Using this simple approximation, we obtain the result that the ground state $E_g$ depends solely on $\alpha_1$. The best approximation exceeds the HF limit by about 15 %. Various tasks, such as resonance scatter, nuclear reactions, and spin-orbit coupling cannot be described; the cross section of the pure potential scatter is also 12 % too low.

Using 14 $\alpha_1$-dependent and 14 $\alpha_2$-dependent functions, we have obtained the HF limit and virtually-excited states (a bound excited state does not exist). The HF wave-function had to be subjected to virtually-excited configurations, i.e., all possible singlet and triplet states. This calculation had to be completed by introducing a further proton (interaction proton) and including all virtual configurations (besides a configuration with three independent nucleons, a configuration of a virtual He³ state). Thus, for low proton energies (slightly above $E_{Th}$), the He³ formation is possible. The exceeding energy can be transferred to the total system and/or to rotations/vibrations of He³. In the same fashion, we have to proceed to the calculations for other nuclei: the configurations of all possible fragments have also to be taken into account. (The cases, corresponding to the O nucleus, are given in listing (18)). In order to keep these considerations short, we now only give a skeleton of the calculation procedures, which are necessary to evaluate the cross sections. When – besides the ground state – all excited states (including virtually-excited states and configurations of fragments) are determined (wave-functions and related energy levels), then Green's function is readily determined by taking the sum over all states. This function contains all coordinates in the configuration space (including the spin), quantum numbers of oscillations, and rotational bands:

$$G(r_n, r'_n) = \sum_{j=0}^{2 \cdot N} \psi^*_j(r'_n) \cdot \psi_j(r_n) \quad (41)$$

The S-matrix is given by:

$$S_{kl} = \int \psi^*_k(r'_n) \cdot G(r_n, r'_n) \cdot \psi_l(r_n) d^3r_1 \ldots d^3r_{2N} \cdot d^3r'_1 \ldots d^3r'_{2N} \quad (42)$$

The transition matrix $T_{kl}$ is defined by all transitions with $k \neq l$:

$$T_{kl} = S_{kl} - \delta_{kl} \quad (43)$$

In order to determine the differential cross section, we need the transition probability. For this purpose, we assume that, before the interaction of the proton with the nucleus, this nucleus is in the ground state. Thus, it might be possible that a proton produces excited states of the nucleus by resonance scatter (inelastic), and a second proton hits the excited nucleus before the transition to the ground state (by emission of a $\gamma$ quantum) has occurred. The second proton would require a lower energy to release either a nucleon or to induce a much higher excited state of the nucleus. However, due to the nuclear cross section, the probability for an inelastic nuclear reaction is very small and would require a very high proton density to yield a noteworthy effect. Therefore, we have calculated the transition probability using the assumption that the occupation probability of the ground state $P_0$ is 1, i.e., $P_0 = 1$ and $P_k = 0$ ($k > 1$). (This is very special case of the Pauli master equation). The differential cross section is obtained by the transition probability divided by the incoming proton flux:

$$\mathrm{dq/d\Omega} = \frac{\text{Transition probability}}{\text{Incoming proton current}} \quad (44)$$

At lower energies, this flux could be calculated by the current given by the Schrödinger equation. To be consistent, we have always used the Dirac equation, since proton energies E > 200 MeV show a significant relativistic effect. With regard to the incoming proton current, we have to point out an important feature:

- The Breit-Wigner formula only considers S states and the incoming current is along the z direction.

- The generalization of this formula by Flügge (1948) includes P states, but the incoming beam is also restricted to the z direction.

Since for our purpose it is necessary to take account for the x/y/z direction by $k_x$, $k_y$, $k_z$ (angular distribution) in the Dirac equation, we have not yet succeeded in obtaining a compact and simple analytical form.

We have already pointed out that the main purpose for calculations with the extended nuclear shell theory incorporate nuclear reaction contributions of protons, neutrons and further small nuclei to the total nuclear cross sections of nuclei discussed in this presentation. We should also mention that the default calculation procedure of nuclear reactions in GEANT4 is an evaporation/cascade model, which has been developed on the basis of statistical thermodynamics.

## Appendix B: Stopping power $S_{sp}$ of secondary protons (online part)

In order to evaluate the transport of the released secondary protons, $He^3$ and $He^4$ ions, we require the stopping power function dE/dz, which is calculated by the following formula (see Ulmer 2007 and Ulmer et al 2009):

$$
\left.
\begin{aligned}
E(z) &= (R_{CSDA} - z) \cdot \sum_{k=1}^{5} A'_k \cdot \exp[-(R_{CSDA} - z)/\beta'_k] \\
dE/dz &= (R_{CSDA} - z) \cdot \sum_{k=1}^{5} A'_k \cdot \beta'^{-1}_k \exp[-(R_{CSDA} - z)/\beta'_k] \\
&\quad - \sum_{k=1}^{5} A'_k \cdot \exp[-(R_{CSDA} - z)/\beta'_k] \\
A'_k &= A_k \cdot (A(water)/Z(water)) \cdot Z \cdot \rho \cdot (75.1/E_I)^{qk}/(A \cdot \rho_w) \\
\beta'_k &= \beta_k \cdot (Z(water) \cdot \rho_w/A(water)) \cdot (75.1/E_I)^{pk} \cdot A/(\rho \cdot Z)
\end{aligned}
\right\} \quad (45)
$$

In Eq. (45) water serves as a reference medium; A(water)/Z(water) is simply 18/10 (Bragg rule), $\rho_w=1$ g/cm$^3$ . With regard to slow neutron cross-section and transport we have used results from Ivanchenko et al 2003 Stankovskiy et al 2007, Zhang and Newhauser 2009, and GEANT4. The parameters of Eq. (45) are stated in Table 5.

The range $R_{CSDA}$ can be calculated with the formula (Ulmer 2007):

$$
R_{CSDA} = \sum_{n=1}^{N} a_n E_0^n \quad (\lim N \Rightarrow \infty) \quad (46)
$$

Formula (46) is only valid for water. If the initial energy $E_0$ satisfies $E_0 < 300$ MeV (therapeutic energies), N = 4 is sufficient (Table 6). For media other than water Eq. (46) has to be replaced by (the restriction to N = 4 remains, see Table 7):

$$
R_{CSDA} = \frac{1}{\rho} \cdot \frac{A}{Z} \cdot \sum_{n=1}^{N} \alpha_n E_I^{pn} E_0^n \quad (\lim N \Rightarrow \infty) \quad (47)
$$

**Table 5.** Parameter values for the inversion Eq. (47) with N = 5, if $E_0$ is in MeV, $E_I$ in eV and $R_{CSDA}$ in cm (dimension of $A_k$: MeV/cm, $\beta_k$: cm). Note: Eq. (45) only requires $A_1,...,A_5$ and $\beta_1,...,\beta_5$.

| $A_1$ | $A_2$ | $A_3$ | $A_4$ | $A_5$ | $\beta_1$ | $\beta_2$ | $\beta_3$ | $\beta_4$ | $\beta_5$ |
|---|---|---|---|---|---|---|---|---|---|
| 99.639 | 25.055 | 8.8075 | 4.19001 | 9.1832 | 0.0975 | 1.24999 | 5.7001 | 10.6501 | 106.727 |

| $P_1$ | $P_2$ | $P_3$ | $P_4$ | $P_5$ | $q_1$ | $q_2$ | $q_3$ | $q_4$ | $q_5$ |
|---|---|---|---|---|---|---|---|---|---|
| -0.1619 | -0.0482 | -0.0778 | 0.0847 | -0.0221 | 0.4525 | 0.195 | 0.2125 | 0.06 | 0.0892 |

**Table 6.** Parameter values for Eq. (46) if $E_0$ is in MeV, $E_I$ in eV and $R_{CSDA}$ in cm.

| $a_1$ | $a_2$ | $a_3$ | $a_4$ |
|---|---|---|---|
| $6.94656 \cdot 10^{-3}$ | $8.13116 \cdot 10^{-4}$ | $-1.21068 \cdot 10^{-6}$ | $1.053 \cdot 10^{-9}$ |

**Table 7.** Parameter values for the inversion Eq. (45) with N = 5, if $E_0$ is in MeV, $E_I$ in eV and $R_{CSDA}$ in cm (dimension of $A_k$: MeV/cm, $\beta_k$: cm). Note: Eq. (47) only requires $A_1,…,A_5$ and $\beta_1,..,\beta_5$.

| $A_1$ | $A_2$ | $A_3$ | $A_4$ | $A_5$ | $\beta_1$ | $\beta_2$ | $\beta_3$ | $\beta_4$ | $\beta_5$ |
|---|---|---|---|---|---|---|---|---|---|
| 99.639 | 25.055 | 8.8075 | 4.19001 | 9.1832 | 0.0975 | 1.24999 | 5.7001 | 10.6501 | 106.727 |
| $P_1$ | $P_2$ | $P_3$ | $P_4$ | $P_5$ | $q_1$ | $q_2$ | $q_3$ | $q_4$ | $q_5$ |
| -0.1619 | -0.0482 | -0.0778 | 0.0847 | -0.0221 | 0.4525 | 0.195 | 0.2125 | 0.06 | 0.0892 |

According to the results of previous sections we have to distinguish between reaction and non-reaction protons. The fluence decrease of non-reaction protons $\Phi_{sp,n}$ is rather similar as in Eq. (1) for primary protons. The related parameters are stated below (details: Ulmer et al 2009). Thus with respect to the stopping power of non-reaction protons $S_{sp,n}$ we can use Eq. (48) and Eq. (49) combined with the aforementioned stated formulas (45 – 47). Since dE/dz in Eq. (45) is based on the framework of CSDA without any energy/range straggling, an appropriate Gaussian convolution has to be applied to reach a poly-chromatic spectral distribution (details: Ulmer et al 2009).

$$\Phi_{sp,n} = (1 - C) \cdot \left( \frac{uq \cdot z}{R_{csda}} \right) \cdot \Phi_0 \right\} (48)$$

$$C = \begin{pmatrix} 0 & (if \ E_0 < E_{Th}) \\ 0.0000008528 & 79 \cdot (E_0 - E_{Th})^2 \ (if \ E_0 \geq E_{Th}) \end{pmatrix} (49)$$

$$z_{shift} = \begin{cases} 0 & (if \ E_0 < E_{res}) \\ \sum_{n=1}^{4} a_n E_{res}^n \cdot [1 - \exp\left(-\frac{(E_0 - E_{res})^2}{5.2^2 \cdot E_{res}^2}\right)] & (if \ E_0 \geq E_{res}) \end{cases} \right\} (50)$$

A tedious task is the determination of the energy transport of the reaction protons $S_{sp,r}$. Thus it is necessary to evaluate the above formulas (45 – 50) along the path of the reaction protons with the corresponding energy by suitable averaging procedures with regard to convolutions; Figures 4 - 5 do not yet provide final information about the contribution of $S_{sp,r}$. According to Figures 1 and 4, the contribution of reaction protons is particular important for E > 100 MeV with increasing energy. We have carried out a statistical analysis of the weights of the proton spectra emerging along the track (Ulmer et al 2009). We now present the calculation formulas for this case. Thus $S_{sp,r}$ is proportional to $\Phi_0 \cdot C$ and a function $F_r$, depending on some further parameters. We use the following definitions and abbreviations:

$$
\left.\begin{aligned}
&z_R = R_{CSDA} / \pi \\
&\tau_s = 0.55411; \; \tau_f = \tau_s - 0.000585437 \cdot (E - E_{res}) \\
&\mathbf{arg\,1} = z / \sqrt{\tau_s^2 + \tau_{in}^2 + (R_{CSDA} / 4\pi)^2} \\
&\mathbf{arg\,2} = (R_{CSDA} - z - \sqrt{2} \cdot \pi \cdot z_{shift}) / \sqrt{\tau_f^2 + (\tau_{straggle} / 7.07)^2 + (R_{CSDA} / \sqrt{3} \cdot 4\pi)^2} \\
&\mathbf{arg} = (R_{CSDA} - 0.5 \cdot z_{shift} \cdot \sqrt{\pi} - z) / (\sqrt{\pi} \cdot z_{shift})
\end{aligned}\right\} \quad (51)
$$

With the help of Eq. (51) we are able to define the calculation procedure of $S_{sp,r}$ by the following formulas, summarized by Eq. (52):

$$
\left.\begin{aligned}
&S_{sp,r} = (E_0 / N_{abs}) \cdot \Phi_0 \cdot [C \cdot F_r + G] \cdot Q^{tot}_{as} \,(medium\,) \cdot A_N^{1/3} / (Q^{tot}_{as} \,(oxygen\,) \cdot A_{oxygen}^{1/3}) \\
&F_r = \varphi \cdot [\varphi_1 + \varphi_2 \cdot \theta] \\
&\varphi = 0.5 \cdot (1 + erf\,(\mathbf{arg})); \quad \varphi_1 = erf\,(\mathbf{arg\,1}); \quad \varphi_2 = erf\,(\mathbf{arg\,2}) \\
&\theta = \begin{pmatrix} e^{-1} - \exp(-(z - z_R)^2 / z_R^2) / (e^{-1} - 1) & (if\; z \le z_R) \\ 1 & (else) \end{pmatrix} \\
&G = (c_1 \cdot z_{shift} \cdot \sqrt{\pi} / R_{CSDA}) \cdot \exp(-(\tfrac{1}{\sqrt{\pi}} R_{CSDA} - z)^2 / z_R^2)
\end{aligned}\right\} \quad (52)
$$

Formulas (51– 52) can only be partially derived, and the adaptation to computed data with the help of the extended nuclear-shell theory is also needed. This can be seen best via computation model M3 (Ulmer et al 200), of which the main contribution consists of the term $I_2$ indicating a proportionality to $[erf(z/\tau) + erf((R_{CSDA} - z)/\tau)]$, if the particles are emerging at surface (i.e., $erf(z/\tau)$ at $z = 0$, whereas the integration boundary $z \to -\infty$ implies the term $[1 + erf((R_{CSDA} - z)/\tau)]$).